\documentclass[a4paper,usenames,dvipsnames,11pt]{article}
\pdfoutput=1

\usepackage{jheppub}
\usepackage{slashed}
\usepackage{mathrsfs,booktabs,multirow,tabularx}
\usepackage{stmaryrd}
\usepackage{xspace}
\usepackage{fancyvrb}
\usepackage[makeroom]{cancel}
\usepackage{amsmath}    
\usepackage{amssymb}    %
\usepackage{graphicx}   
\usepackage{verbatim}   
\usepackage{lscape}
\usepackage{subfig}
\usepackage{listings}
\usepackage{pifont}
\usepackage{mathtools}

\def\beq{\begin{equation}}
\def\beqn{\begin{eqnarray}}
\def\eeq{\end{equation}}
\def\eeqn{\end{eqnarray}}

\def\binomial#1#2{
\left(\!
\begin{array}{c}
#1\\
#2
\end{array}
\!\right)
}
\def\stirlingSo#1#2{
\left[\!
\begin{array}{c}
#1\\
#2
\end{array}
\!\right]
}
\def\stirlingSt#1#2{
\left\{\!
\begin{array}{c}
#1\\
#2
\end{array}
\!\right\}
}

\newcommand{\oPi}{\overline{\Pi}}
\newcommand{\opi}{\bar{\pi}}
\newcommand{\brho}{\bar{\rho}}

\title{Some identities which involve Stirling numbers}

\author[a]{Stefano Frixione}
\affiliation[a]{INFN, Sezione di Genova, Via Dodecaneso 33, I-16146, 
Genoa, Italy}

\emailAdd{Stefano.Frixione@cern.ch}

\abstract{During the course of an ongoing work on the small-$x$ behaviour 
of parton distribution functions, some identities have been found which 
involve Stirling numbers of the first and the second kind, as well as 
binomial coefficients. Without any claim of originality I report them 
in this note.}


\begin{document}
\maketitle
\flushbottom

\section{Notation\label{sec:not}}
Unless specified otherwise, all numbers which appear in this
note are integers. By
\beq
\binomial{n}{k},\;\;\;\;\;\;\;\;
\stirlingSo{n}{k},\;\;\;\;\;\;\;\;
\stirlingSt{n}{k},
\label{defs}
\eeq
I denote the binomial coefficient (i.e.~the coefficient of $x^k$ in
the series expansion of \mbox{$(1+x)^n$}), the unsigned Stirling numbers
of the first kind (i.e.~the coefficient of $x^k$ in the series expansion of 
the rising factorial \mbox{$x^{(n)}$}; equivalently, the number of
permutations of $n$ elements with $k$ disjoint cycles), and the Stirling 
numbers of the second kind (i.e.~the number of ways to partition a set
of $n$ elements into $k$ non-empty subsets; equivalently, the coefficient 
of the falling factorial \mbox{$x_{(k)}$} in the expansion of $x^n$ in
terms of such factorials), respectively. I remind the 
reader that the (signed) Stirling numbers of the first kind, $s(n,k)$, 
can immediately be obtained from the unsigned ones by parity:
\beq
s(n,k)=(-)^{n-k}\,\stirlingSo{n}{k}\,.
\label{sS1}
\eeq
By
\beq
\Pi^{(p)}\,,\;\;\;\;\;\;\;\;
\oPi^{(p)}\,,
\label{intpart}
\eeq
I denote the set of the unordered and ordered\footnote{Since this can be 
confusing, I stress explicitly that the elements of an unordered partition
$\pi^{(p)}_a$ {\em may} have a {\em conventional} ordering (eq.~(\ref{p2})), 
but such an ordering must not be relevant to the result of any function of 
$\pi^{(p)}_a$. Conversely, the ordering of the elements of an ordered 
partition $\opi^{(p)}_a$ generally does matter for the result of a function of 
$\opi^{(p)}_a$ -- therefore, all possible orderings must be considered.}
integer partitions of $p$, respectively. Then:
\beqn
\pi^{(p)}_a\in\Pi^{(p)}\;\;&\Longrightarrow&\;\;
\pi^{(p)}_a=\big\{l_1,l_2,\ldots l_{m_a}\big\}\,,\;\;\;\;
\sum_{m=1}^{m_a}l_{m}=p\,,
\label{p1}
\\*
&&\;\;1\le m_a\le p\,,\;\;\;\;
1\le l_1\le l_2\le\ldots l_{m_a}\le p\,,
\label{p2}
\eeqn
and:
\beqn
\!\!\!\!\opi^{(p)}_a\in\oPi^{(p)}\;\;&\Longrightarrow&\;\;
\opi^{(p)}_a=\big\{l_1,l_2,\ldots l_{m_a}\big\}\,,\;\;\;\;
\sum_{m=1}^{m_a}l_{m}=p\,,
\label{op1}
\\*
&&\;\;1\le m_a\le p\,,\;\;\;\;
1\le l_m\le p\;\;\forall\,m\,.
\label{op2}
\eeqn
The number of unordered and ordered integer partitions of $p$ is
equal to
\beq
\#\left(\Pi^{(p)}\right)=\rho(p)\,,\;\;\;\;\;\;\;\;
\#\left(\oPi^{(p)}\right)=\brho(p)\equiv 
\left(1-\delta_{0p}\right)\,2^{p-1}+\delta_{0p}\,,
\eeq
respectively (thus, $1\le a\le\rho(p)$ for $\pi^{(p)}_a$, 
and $1\le a\le\brho(p)$ for $\opi^{(p)}_a$), with:
\beq
\sum_{p=0}^\infty \rho(p)x^p=\prod_{j=1}^\infty\frac{1}{1-x^j}=
1+x+2x^2+3x^3+5x^4+7x^5+11x^6+\ldots\,.
\eeq
By construction, any ordered partition is the (possibly trivial)
permutation of an unordered partition. The map $\oPi^{(p)}\to\Pi^{(p)}$
is surjective and, in keeping with this, $\brho(p)\ge\rho(p)$ $\forall\,p$.

I also denote by
\beq
\oPi^{(p,q)}
\eeq
the set of the ordered partitions of $p$ that contain exactly $q$ 
elements\footnote{Henceforth, the elements $l_m\in\opi^{(p,q)}_a$ may 
also be labelled with $0\le m\le q-1$.\label{ft:label}}:
\beqn
\!\!\!\!\opi^{(p,q)}_a\in\oPi^{(p,q)}\;\;&\Longrightarrow&\;\;
\opi^{(p,q)}_a=\big\{l_1,l_2,\ldots l_q\big\}\,,\;\;\;\;
\sum_{m=1}^{q}l_{m}=p\,,
\label{opq1}
\\*
&&\;\;
0\le l_m\le p\;\;\forall\,m\,.
\label{opq2}
\eeqn
There are
\beq
\#\left(\oPi^{(p,q)}\right)=\brho(p,q)\equiv 
\binomial{p+q-1}{q-1}
\label{brhodef}
\eeq
of such partitions (therefore, $1\le a\le\brho(p,q)$ for $\opi^{(p,q)}_a$). 
Note (eq.~(\ref{opq2})) that the constraint that there be exactly $q$ 
elements in any partition implies that such elements (at variance with 
those of both $\pi^{(p)}_a$ and $\opi^{(p)}_a$) may be equal to zero.

Finally, $\Pi^{(p)}$ can be mapped bijectively onto a set $K^{(p)}$
such that
\beqn
k^{(p)}_a\in K^{(p)}\;\;&\Longrightarrow&\;\;
k^{(p)}_a=\big\{k_1,k_2,\ldots k_p\big\}\,,\;\;\;\;
\sum_{m=1}^{p}mk_{m}=p\,,
\label{kq1}
\\*
&&\;\;
0\le k_m\le p\;\;\forall\,m\,.
\label{kq2}
\eeqn
Since the map is bijective, $\Pi^{(p)}$ and $K^{(p)}$ have the same
number of elements, and any $k^{(p)}_a$ can be seen as a representation
of some $\pi^{(p)}_a$.

\section{Stirling numbers of the first kind\label{sec:S1}}
For any $p$ and $q$ one finds the following identity:
\beqn
\stirlingSo{q+p}{q}=
\sum_{\{l_m\}_{m=1}^{m_a}\in\oPi^{(p)}}\,
\frac{\Gamma\left(1+q+p\right)}
{\Gamma(q-m_a+1)\prod_{m=1}^{m_a}\left(\sum_{k=1}^m (l_k+1)\right)}\,,
\phantom{aaa}
\label{S1id1}
\eeqn
which thus gives the Stirling numbers of the first kind as a sum
of contributions stemming from {\em ordered integer partitions}; note 
that all such contributions with $m_a>q$ vanish, owing to the $\Gamma$ 
function in the denominator. I point out that there exists another
expression~\cite{malenfant2011finiteclosedformexpressionspartition} where 
the Stirling numbers of the first kind are given as a sum over 
{\em unordered integer partitions}, in the representation
of eq.~(\ref{kq1}), namely:
\beq
s(n,n-p)=\frac{1}{(n-p-1)!}\sum_{\{k_m\}_{m=1}^{p}\in K^{(p)}}
\frac{(-)^k(n+k-1)!}
{k_1!k_2!\ldots k_p!2!^{k_1}3!^{k_2}\ldots (p+1)!^{k_p}}\,,
\label{WKs1}
\eeq
where $k=k_1+k_2+\ldots +k_p$. Although $\oPi^{(p)}$ can be mapped onto 
$K^{(p)}$ (via $\Pi^{(p)}$), in general the sequences of summands
stemming from eq.~(\ref{S1id1}) and eq.~(\ref{WKs1}) are different
from one another -- among other things, only the latter features
both positive and negative contributions, while with the former
one simply obtains the signed Stirling numbers of the first kind
by means of eq.~(\ref{sS1}) from an all-positive sequence.
For example, one has
\beq
\stirlingSo{7}{3}=1624=504+420+280+210+90+72+48
\eeq
by using eq.~(\ref{S1id1}), and
\beq
\stirlingSo{7}{3}=1624=-21+280+420-3780+4725
\eeq
by using eq.~(\ref{WKs1}). For comparison, the sequences emerging
from the standard double-sum and Vieta's formulae for the Stirling
numbers of the first kind are
\beqn
\stirlingSo{7}{3}&=&1624=-3465-1540+49280-198+25344-144342
\nonumber
\\*&&\phantom{=1624}
-15/2+2880-98415/2+122880\,,
\\*
\stirlingSo{7}{3}&=&1624=24+30+36+40+48+60+60+72+90+120+120
\nonumber
\\*&&\phantom{=1624}
+144+180+240+360\,,
\eeqn
respectively.

\section{Stirling numbers of the second kind\label{sec:S2}}
Other identities relate expressions involving binomial coefficients 
on one side, and Stirling numbers of the second kind on the other side. 
The basic quantities from which these identities originate are 
defined as follows (see footnote~\ref{ft:label} for labelling conventions):
\beqn
c_l\!\left(n,q;\big\{n_i\big\}\right)\!&=&\!
\sum_{\{j_i\}_{i=0}^q\in\oPi^{(l,q+1)}}
\left(\prod_{i=0}^{q} (i+1)^{n_i-j_i}\binomial{n_i}{j_i}\right)
\label{cjcffmod}
\\*\!&\equiv&\!
\sum_{\{j_i\}_{i=0}^q\in\oPi^{(l,q+1)}}
\frac{1^{n_0}}{1^{j_0}}\binomial{n_0}{j_0}
\frac{2^{n_1}}{2^{j_1}}\binomial{n_1}{j_1}
\frac{3^{n_2}}{3^{j_2}}\binomial{n_2}{j_2}\ldots
\frac{(q+1)^{n_q}}{(q+1)^{j_q}}\binomial{n_q}{j_q},\phantom{aaaa}
\label{cjcff0b}
\eeqn
for any $n\ge 1$, $q\ge 1$, and with $0\le l\le n$. I shall then consider
(note: $n=\sum_{i=0}^q n_i$)
\beqn
\bar{c}_l\!\left(n,q;w\right)&=&
\sum_{\{n_i\}_{i=0}^q\in\oPi^{(n,q+1)}}
\!\!\!w\!\left(\big\{n_i,m_i\big\}\right)
c_l\!\left(n,q;\big\{n_i\big\}\right)\,,
\label{bcsol0Wmod}
\eeqn
with either
\beq
w\!\left(\big\{n_i,m_i\big\}\right)=1
\label{weq1}
\eeq
or
\beq
w\!\left(\big\{n_i,m_i\big\}\right)=
\prod_{i=1}^\lambda (1+n_{\alpha_i})^{(m_i)}\,,\;\;\;\;\;\;\;\;
1\le\lambda\le q+1\,.
\label{weqRF}
\eeq
In eq.~(\ref{weqRF}) the $\lambda$ integers $m_i$ are given, and 
$\alpha_i$ are such that
\beq
0\le\alpha_i\le q\,,\;\;\;\;\;\;\;\;
\alpha_i\ne\alpha_j\;\;\;\forall\,i\ne j\,.
\eeq
I have used a standard notation for the rising factorial of
\mbox{$1+n_{\alpha_i}$}, which however will be shortened henceforth 
as follows:
\beq
\alpha^{(m)}:=
(1+n_\alpha)^{(m)}\equiv
\frac{\Gamma(n_\alpha+m+1)}{\Gamma(n_\alpha+1)}\,.
\label{atomshort2}
\eeq
The first identity stems from the choice in eq.~(\ref{weq1}),
and reads thus:
\beqn
\bar{c}_l\!\left(n,q;1\right)&=&
\binomial{n+q}{l}\stirlingSt{n-l+q+1}{q+1}\,,
\label{bcsolmod}
\eeqn
or, more explicitly:
\beq
\!\!\!\!\sum_{\{n_i\}_{i=0}^q\in\oPi^{(n,q+1)}}
\sum_{\{j_i\}_{i=0}^q\in\oPi^{(l,q+1)}}
\left(\prod_{i=0}^{q} (i+1)^{n_i-j_i}\binomial{n_i}{j_i}\right)=
\binomial{n+q}{l}\stirlingSt{n-l+q+1}{q+1}.
\label{bcsolmod2}
\eeq
When $l=0$ eq.~(\ref{bcsolmod2}) coincides with the expression of
the Stirling numbers of the second kind quoted in Wikipedia as
coming from the NIST Handbook of Mathematical Functions, namely:
\beq
\stirlingSt{n}{k}=\sum_{c_1,\ldots c_k=0}^{c_1+\ldots c_k=n-k}
1^{c_1}2^{c_2}\ldots k^{c_k}\,.
\eeq
Conversely, for any $l\ge 1$ neither the sequences 
$\{\bar{c}_l\!\left(n,q;1\right)\}_n$ at fixed $q$ nor the sequences
$\{\bar{c}_l\!\left(n,q;1\right)\}_q$ at fixed $n$, in spite of their
simple expression in eq.~(\ref{bcsolmod2}), are included in the OEIS
(with the possible exceptions of those stemming from the lowest values 
of $q$ and of $n$ with $n\ge l$, respectively). 

Turning now to eq.~(\ref{weqRF}), I begin with the case where there
is a single rising factorial (i.e.~$\lambda=1$). One obtains (bearing
in mind the shorthand notation of eq.~(\ref{atomshort2})):
\beqn
\bar{c}_l\!\left(n,q;\alpha^{(m)}\right)&=&
m!\binomial{n+q+m}{l}\,\hat{c}(n-l,q,m,\alpha)\,,
\label{albega1b2}
\\*
\hat{c}(j,q,m,\alpha)&=&
\sum_{s=0}^j \binomial{j-s+m-1}{j-s}
\big(\alpha+1\big)^{j-s}
\stirlingSt{s+q+1}{q+1}.\phantom{aaaa}
\label{hcdef}
\eeqn
This result agrees with eq.~(\ref{bcsolmod}) when $m=0$, if one
uses the convention \mbox{$(-1\;\;0)=1$}.
Equations~(\ref{albega1b2}) and~(\ref{hcdef}) thus express \mbox{$2(q+1)$} 
nested sums (or, equivalently, two sums over the integer partitions of $n$ and 
$l$ that contain exactly \mbox{$q+1$} elements) of binomial coefficients, 
times power-like prefactors and a single rising factorial, in terms of a 
single sum whose summands feature Stirling numbers of the second kind. 
Because of this single sum such an expression is more complicated
than that in eq.~(\ref{bcsolmod}), which is a consequence
of the presence of the rising factorial in $\bar{c}_l$.

Finally, with $\lambda>1$ one obtains:
\beqn
\!\!\!\!\!\bar{c}_l\!
\left(n,q;\prod_{i=1}^\lambda\alpha_i^{(m_i)}\right)&=&
\sum_{k_1=1}^{m_1}
\sum_{k_2=m_2}^{m-k_1}\ldots\sum_{k_\lambda=m_\lambda}^{m-k_1}
\delta\big(k_2+\ldots +k_\lambda-(m-k_1)\big)
\nonumber
\\*&&\phantom{aa}\times
\bar{c}_l\!\left(n,q;\alpha_1^{(k_1)}\right)
\frac{r_l(m;k_1)}{k_1!}\,\left(\prod_{i=1}^\lambda m_i!\right)\,
\left[\prod_{i=2}^\lambda \binomial{k_i-1}{k_i-m_i}\right]\,
\nonumber
\\*&&\phantom{aa}\times
\frac{(-)^{m-m_1}(1+\alpha_1)^{m-m_1}
\Big[\prod_{i=2}^\lambda (1+\alpha_i)^{k_i-m_i}\Big]}
{\Big[\prod_{i=2}^\lambda(\alpha_i-\alpha_1)^{k_i}\Big]}
\nonumber
\\*&+&
\Big(1\to 2\to 3\to\ldots \lambda\to 1\Big)
\big[\lambda-1~{\rm times}\big]\,,
\label{albegawl}
\eeqn
where
\beq
m=\sum_{i=1}^\lambda m_i\,,
\eeq
and
\beq
r_l(a;b)=\binomial{n+q+a}{l}\Bigg/\binomial{n+q+b}{l}\,.
\label{rm1m2mod}
\eeq
The last line of eq.~(\ref{albegawl}) compactly denotes $(\lambda-1)$
contributions functionally identical to that which appears in the first 
three lines on the r.h.s.~bar for the cyclic permutations over the indices
\mbox{$\{1,\ldots\lambda\}$}. For example, the case $\lambda=2$ reads 
as follows:
\beqn
\!\!\!\!\!\bar{c}_l\!\left(n,q;\alpha_1^{(m_1)}\alpha_2^{(m_2)}\right)&=&
\sum_{k_1=1}^{m_1}
\bar{c}_l\!\left(n,q;\alpha_1^{(k_1)}\right)
r_l(m;k_1)\,\frac{m_1!m_2!}{k_1!}\,
\binomial{k_2-1}{k_2-m_2}
\nonumber
\\*&&\phantom{aaaaaaaaa}\times
\left.
\frac{(-)^{m_2}(1+\alpha_1)^{m_2}(1+\alpha_2)^{k_2-m_2}}
{(\alpha_2-\alpha_1)^{k_2}}
\right|_{k_2=m-k_1}
\nonumber
\\*&+&
\sum_{k_2=1}^{m_2}
\bar{c}_l\!\left(n,q;\alpha_2^{(k_2)}\right)
r_l(m;k_2)\,\frac{m_1!m_2!}{k_2!}\,
\binomial{k_1-1}{k_1-m_1}
\nonumber
\\*&&\phantom{aaaaaaaaa}\times
\left.
\frac{(-)^{m_1}(1+\alpha_2)^{m_1}(1+\alpha_1)^{k_1-m_1}}
{(\alpha_1-\alpha_2)^{k_1}}
\right|_{k_1=m-k_2}\,.\phantom{aaa}
\label{albegaw2}
\eeqn
Equation~(\ref{albegaw2}) is trivial as far as the summation indices 
(bar the outermost one) are concerned. In general, and by taking the
first line as a representative case, in eq.~(\ref{albegawl}) the sums 
over the indices \mbox{$\{k_2\ldots k_\lambda\}$} are equivalent, owing 
to the Kronecker $\delta$, to summing over the integer partitions of 
$(m-k_1)$ that contain exactly \mbox{$\lambda-1$ elements}, i.e.
\beq
\Big\{k_2\,,\ldots k_\lambda\Big\}\in\oPi^{(m-k_1,\lambda-1)}\,.
\eeq
These indices are subject to the further constraints $k_i\ge m_i$, 
which are in any case enforced  by the binomial coefficients.
Crucially, eq.~(\ref{albegawl}) also shows that the single rising factorial
result of eqs.~(\ref{albega1b2}) and~(\ref{hcdef}) is the key quantity
for expressing $\bar{c}_l\!\left(n,q;w\right)$ in terms of the
Stirling numbers of the second kind also in the case of multiple
rising factorials. Finally, note the following significant simplification
in eq.~(\ref{albegawl})
\beq
\bar{c}_l\!\left(n,q;\alpha_i^{(k_i)}\right)\frac{r_l(m;k_i)}{k_i!}=
\binomial{n+q+m}{l}
\hat{c}(n-l,q,k_i,\alpha_i)
\eeq
which occurs for any $i$. It thus becomes manifest that eq.~(\ref{albegawl}) 
coincides with eq.~(\ref{albega1b2}) by setting $m_i=\alpha_i=0$ for all
$i\ge 2$ and, in keeping with that, $\lambda=1$.

In summary, the l.h.s.~of eq.~(\ref{albegawl}) features $2(q+1)$ nested
sums (or two sums over integer partitions with exactly $q+1$ elements)
of $(q+1)$ binomial coefficients times \mbox{$(q+1)$} power-like coefficients,
times $\lambda$ rising factorials. It is expressed on the r.h.s.~of that
equation by means of $\lambda+1$ nested sum (or two sums over single indices, 
plus a sum over integer partitions with exactly $\lambda-1$ elements), 
cyclically symmetrised, of a single instance of Stirling numbers of the
second kind, times $\lambda$ rational coefficients times $\lambda$
binomial coefficients.

In the context of ref.~\cite{Bonvini:2025xxx} it is of interest to
study the $n\to\infty$ properties of the quantities on the l.h.s.~of
eq.~(\ref{albegawl}) for generic, but finite, values of $q$, $\lambda$,
$\alpha_i$, and $m_i$. This can be done by using the representation
on the r.h.s.~of eq.~(\ref{albegawl}), where the only dependence upon $n$ 
is that of the coefficients $\bar{c}_l(n,q;\alpha_i^{(k_i)})$.
These, in turn, can be evaluated by means of eqs.~(\ref{albega1b2}) 
and~(\ref{hcdef}). However, the fact that the upper limit of the sum
in the latter equation is equal to $n-l$ is inconvenient when one considers 
the $n\to\infty$ limit, and finding an alternative expression is therefore 
useful. This can be achieved by starting from the observation that, at 
given $q$, $m$, and $\alpha$, the coefficients $\hat{c}(j,q,m,\alpha)$
can formally be expressed in terms of a generating function
\beq
\widehat{C}(t;q,m,\alpha)=\sum_{j=0}^\infty
\hat{c}(j,q,m,\alpha)\,t^j\,,
\eeq
for which one finds:
\beq
\widehat{C}(t;q,m,\alpha)=
\left[\Big(1-(\alpha+1)t\Big)^m
\prod_{i=0}^{q}\Big(1-(i+1)t\Big)\right]^{-1}\,.
\eeq
One then rewrites this as follows:
\beqn
\widehat{C}(t;q,m,\alpha)&=&
\exp\left(f(t;\alpha+1,m)+\sum_{i=0}^q f(t;i+1,1)\right),
\\*
f(t;a,b)&=&-b\log(1-at)\,,
\eeqn
and then exploits the expression of the generating function of the
complete exponential Bell polynomials $B_j$, to obtain:
\beq
\hat{c}(j,q,m,\alpha)=\frac{1}{j!}\,
B_j\left(\left\{\frac{d^p}{dt^p}\left.
\left(f(t;\alpha+1,m)+\sum_{i=0}^q f(t;i+1,1)\right)\right|_{t=0}
\right\}_{p\ge 1}\right).
\label{hcwBj}
\eeq
Then one computes
\beq
\left.\frac{d^p}{dt^p}\,f(t;a,b)\right|_{t=0}=(p-1)!\,b\,a^p\,,
\eeq
whence
\beq
B_j\left(\left\{(p-1)!\,b\,a^p\right\}\right)=
a^j\,\frac{\Gamma(b+j)}{\Gamma(b)}
\;\;\;\stackrel{b\to 1}{\longrightarrow}\;\;\;
a^j\,j!\,.
\label{Bellj0}
\eeq
After some algebra, one finally finds:
\beqn
\hat{c}(j,q,m,\alpha)&=&
(-)^{q-\alpha}\,\binomial{j+m}{m}
\binomial{q}{\alpha}\frac{(1+\alpha)^{q+j}}{q!}
\label{Bjwgt1}
\\*&+&
\sum_{\stackrel{i\ne\alpha}{i=0}}^q
(-)^{q-i}\binomial{q}{i}\frac{(1+i)^q}{q!\,(i-\alpha)^m}\,
\Bigg((1+i)^{m+j}
\nonumber
\\*&&\phantom{aaa}+
m\binomial{j+m}{m}(1+\alpha)^j
\sum_{k=1}^m(-)^k\binomial{m-1}{k-1}\frac{(1+\alpha)^k(1+i)^{m-k}}{j+k}
\Bigg).
\nonumber
\eeqn
Equation~(\ref{Bjwgt1}) can be used instead of eq.~(\ref{hcdef});
in the $n\to\infty$ limit, it only features sums with a finite
number of terms. It also gives one identities alternative to those
in eq.~(\ref{albegawl}), where now the r.h.s.~does not feature
the Stirling numbers of the second kind any longer (and is different
w.r.t.~the expression one would obtain by replacing the Stirling
numbers in eq.~(\ref{hcdef}) with their usual single-sum representation).

What is more interesting in the present context is that the extreme 
simplicity of the $b=1$ result in eq.~(\ref{Bellj0}) leads one to
the following compact expression:
\beqn
&&B_j\left(\left\{\frac{d^p}{dt^p}\left.
\left(\sum_{i=0}^q f(t;i+1,1)\right)\right|_{t=0}
\right\}_{p\ge 1}\right)
\\*&&\phantom{aaaaaa}
=B_j\Big(\big\{(p-1)!\left(1+2^p+3^p+\ldots 
(q+1)^p\right)\big\}_{p\ge 1}\Big)
\\*&&\phantom{aaaaaa}
=\frac{j!}{q!}\,
\sum_{i=0}^q (-)^{q-i}\binomial{q}{i}(i+1)^{q+j}\,,
\label{Bellmeq0}
\eeqn
having repeatedly used the binomial-like summation formula of the
complete Bell polynomials. It is then easy to see that eq.~(\ref{Bellmeq0})
is essentially the single-sum expression of the Stirling numbers
of the second kind, whence:
\beq
\stirlingSt{j+q+1}{q+1}=
\frac{1}{j!}\,B_j\Big(\big\{(p-1)!\left(1+2^p+3^p+\ldots 
(q+1)^p\right)\big\}_{p\ge 1}\Big).
\label{SS2Bell}
\eeq
A minor drawback of this result is that the arguments of the complete Bell 
polynomials depend on $q$ (whereas the usual expression for the Stirling 
numbers of the second kind in terms of the {\em incomplete} exponential 
Bell polynomials features a sequence of $1$'s as arguments of the latter);
this is unavoidable, since the complete Bell polynomials form a
one-index sequence.

Finally, I point out that the coefficients of the $(i+1)^j$ terms in
eq.~(\ref{Bellmeq0}) are known as coefficients of the Sidi 
polynomials~\cite{Sidi_2003}. The first few of them read:
\beqn
\Big\{(-)^{q-i}\binomial{q}{i}(i+1)^q\Big\}_{i=0}^q
&=&\big\{-\!1,2\big\}\;\;\;q=1
\\*
&=&\big\{1,-8,9\big\}\;\;\;q=2\phantom{\Big\{}
\\*
&=&\big\{-\!1,24,-81,64\big\}\;\;\;q=3\phantom{\Big\{}
\\*
&=&\big\{1,-64,486,-1024,625\big\}\;\;\;q=4\phantom{\Big\{}
\eeqn
and so forth. These coefficients admit an exponential generating function:
\beq
G(x,y)=-\frac{W\left(-xye^{-x}\right)}
{xy\left(1+W\left(-xye^{-x}\right)\right)}\,,
\label{WEGF}
\eeq
where $W$ denotes the Lambert $W$ function. Therefore:
\beq
\stirlingSt{n}{k}=\frac{1}{(k-1)!}\,
\sum_{i=0}^{k-1}\,\frac{(i+1)^{n-k}}{i!}
\left.\frac{\partial^i\partial^{k-1}}
{\partial y^i\partial x^{k-1}}\,G(x,y)\right|_{(x,y)=(0,0)}\,.
\label{SS2Sidi}
\eeq

\section{Conclusions\label{sec:concl}}
As a by-product of an ongoing work~\cite{Bonvini:2025xxx}, I have
found some identities which involve binomial coefficients and
Stirling numbers of the first and the second kind -- see
eqs.~(\ref{S1id1}), (\ref{bcsolmod2}), (\ref{albega1b2}), (\ref{albegawl}), 
(\ref{Bjwgt1}), (\ref{SS2Bell}), and~(\ref{SS2Sidi}). In view of the fact 
that integer sequences are rarely relevant to high-energy physics, I have 
reported these identities here should someone find them useful (even 
accidentally, as was my case). Although I have been unable to find them
anywhere, I do not claim that they are unknown or unpublished, and 
in fact I will be most grateful to whomever will point me to any
appropriate references.

{\bf Acknowledgements:}
I am happy to thank my collaborators on ref.~\cite{Bonvini:2025xxx}, 
Marco~Bonvini and Giovanni~Stagnitto, for their work on that project,
and for valuable comments on this manuscript.

\phantomsection
\addcontentsline{toc}{section}{References}
\bibliographystyle{JHEP}
\bibliography{strid}

\end{document}